# (3-Aminopropyl)trimethoxysilane Surface Passivation Improves Perovskite Solar Cell Performance by Reducing Surface Recombination Velocity

Yangwei Shi,[1,2] Esteban Rojas-Gatjens,[3] Jian Wang,[1] Justin Pothoof,[1] Rajiv Giridharagopal,[1] Kevin Ho,[1] Fangyuan Jiang,[1] Margherita Taddei,[1] Zhaoqing Yang,[1] Carlos Silva-Acuña,[3,4,5] and David S. Ginger[1]*

1. Department of Chemistry, University of Washington, Seattle, WA 98195, USA
2. Molecular Engineering & Sciences Institute, University of Washington, Seattle, WA 98195, USA
3. School of Chemistry and Biochemistry, Georgia Institute of Technology, Atlanta, GA 30332, USA
4. School of Physics, Georgia Institute of Technology, Atlanta, GA 30332, USA
5. School of Materials Science and Engineering, Georgia Institute of Technology, Atlanta, GA 30332, USA

*Corresponding author: dginger@uw.edu

**Abstract**

We demonstrate reduced surface recombination velocity (SRV) and enhanced power-conversion efficiency (PCE) in mixed-cation mixed-halide perovskite solar cells by using (3-aminopropyl)trimethoxysilane (APTMS) as a surface passivator. We show the APTMS serves to passivate defects at the perovskite surface, while also decoupling the perovskite from detrimental interactions at the $C_{60}$ interface. We measure a SRV of ~125 ± 14 cm/s, and a concomitant increase of ~100 meV in quasi-Fermi level splitting in passivated devices compared to the controls. We use time-resolved photoluminescence and excitation-correlation photoluminescence spectroscopy to show that APTMS passivation effectively suppresses non-radiative recombination. We show that APTMS improves both the fill factor and open-circuit voltage ($V_{OC}$), increasing $V_{OC}$ from 1.03 V for control devices to 1.09 V for APTMS-passivated devices, which leads to PCE increasing from 15.90% to 18.03%. We attribute enhanced performance to reduced defect density or suppressed nonradiative recombination and low SRV at the perovskite/transporting layers interface.

**TOC graphic**

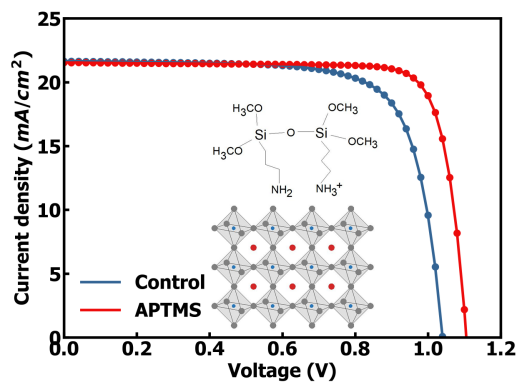

**Introduction**

Halide perovskites are promising semiconductors for use in applications ranging from solar photovoltaics,[1-3] to conventional electroluminescence,[4,5] and even as non-linear optical elements or quantum light sources.[6,7] In photovoltaic applications, the record power-conversion efficiency (PCE) has surged to 25.7% for a single junction perovskite solar cells.[8] Despite this rapid progress, the best efficiencies being obtained today still lag behind the theoretical detailed balance limit of 32.3% (for a ~1.48 eV bandgap absorber), which is largely due to a deficit in the open-circuit voltage ($V_{OC}$) that results from non-radiative recombination in the bulk of perovskite films and also at the interfaces between perovskite and transport layers.[9-13]

Previous studies have shown that surface recombination at the surface/interface of the perovskite layer in solar cells plays an important role on the device performance.[14-16] A small surface recombination velocity (SRV) enables a higher value of $V_{OC}$ and FF.[15] Many different schemes have been proposed for surface passivation. Of these, (3-aminopropyl)trimethoxysilane (APTMS) exhibits many characteristics of an ideal surface passivator: (i) it is effective at passivating a large number of perovskite compositions,[17] (ii) it is commercially available at large scales, (iii) it can be deposited from solution or vapor phases, and (iv) it is polymerizable, which should result in more stable passivation layers. Jariwala *et al.* previously demonstrated an effective APTMS passivation strategy for perovskites on glass that reduces the SRV down to 30 cm/s (average), through which a quasi-Fermi level splitting (QFLS) of 1.31 eV can be achieved for a 1.63 eV bandgap mixed-cation mixed-halide perovskite, approaching its theoretical limit of 1.35 eV.[17] However, they did not examine the effect of APTMS passivation on working solar cells, and because thicker APTMS layers can be insulating, it has remained unclear if these improvements in SRV could be translated into actual solar cell performance.

In this work, we show that, by optimizing the APTMS deposition conditions, the improvements in SRV can be translated into improvements in $V_{OC}$, fill factor (FF), and efficiency of working solar cells. We use a methylammonium (MA)-free mixed-cation mixed-halide perovskite $FA_{0.83}Cs_{0.17}Pb(I_{0.85}Br_{0.15})_3$,[18] denoted hereafter as Cs17Br15, to investigate the APTMS passivation in perovskite solar cells. By using photoluminescence quantum yield (PLQY) measurements and QFLS analysis on partial device stacks and complete perovskite solar cells, we show that the QFLS in passivated devices increases to 1.23 eV, which enhances $V_{OC}$ up to 1.11 V compared to values of 1.10 eV and 1.04 V respectively for the unpassivated control devices.

**Results and discussions**

Figure 1a shows the schematic of the p-i-n structure of the perovskite solar cells used in this study. We use $FA_{0.83}Cs_{0.17}Pb(I_{0.85}Br_{0.15})_3$ (where FA = formamidinium), which has a bandgap of 1.63 eV. We choose this composition since it has a similar bandgap to the archetypal $MAPbI_3$ but exhibits a higher thermal stability.[19] For the hole-transport layer (HTL), we choose a self-assembled monolayer (SAMs) of [2-(3,6-dimethoxy-9H-carbazol-9-yl)ethyl]phosphonic acid (MeO-2PACz) for its good compatibility and easy processability.[20] Details regarding the APTMS layer deposition and perovskite solar cell fabrication can be found in Supporting Information[17]

Figure 1c shows the PL lifetime of the control film and APTMS-passivated film on glass substrates. We obtain a longer PL lifetime (1.2 ± 0.2 μs) for passivated perovskite films in comparison to the control perovskite films (50 ± 8 ns). We extract the SRV from the time-resolved photoluminescence (TRPL) data using the approximations described in our previous work[15,17] (see also SI). From this approach, we conservatively estimate the average SRV after APTMS treatment is reduced to ~ 50 cm/s or lower, consistent with our previous work.[17] Labile surface passivators such as tri-*n*-octylphosphine oxide, and alkanethiols, are difficult to incorporate into devices because, while they have been shown to effectively coordinate dangling bonds on undercoordinated $Pb^{2+}$ sites at the perovskite surface (e.g. halide vacancies),[21,13] they are also easily washed away or pumped off under high vacuum. However, Figure 1d shows that, because APTMS hydrolyzes to form a robust film cross-linked by Si-O-Si bonds[22], APTMS-treated films are able to retain their long PL lifetimes and low SRV values even after washing with solvents such as chlorobenzene (CB) or annealing at 100 °C and high vacuum (Figure 1d and Figure S5) (See also SI). Furthermore, we measure the SRV in partial solar cell stacks without a silver top electrode (ITO/MeO-2PACz/Cs17Br15/APTMS/C60) to be 125 cm/s ± 14 cm/s.

We mapped the optoelectronic properties of APTMS-passivated films by fluorescent lifetime imaging microscopy (FLIM) to assess the homogeneity of passivation over the sample surface.[23] Figure S2a and b show that the control film is much dimmer in comparison to the APTMS-passivated film, while the APTMS -passivated film appears much brighter. Previous reports have shown that additives or processing agents may enhance the PL of perovskite film via the formation of low-dimensional wide-gap perovskites at the surface.[24-26] We performed XRD to examine this possibility but we do not observe any additional diffraction peaks at low angle (Figure S3a) that are associated with the formation of low-dimensional perovskites. UV-vis absorption and steady-state PL measurements show no evidence of a low-dimensional phase emerging in APTMS passivated films (Figure S3b). We thus conclude that APTMS passivation has a negligible influence on the bulk perovskite structure and no (detectable) low-dimensional perovskite formed after surface passivation.

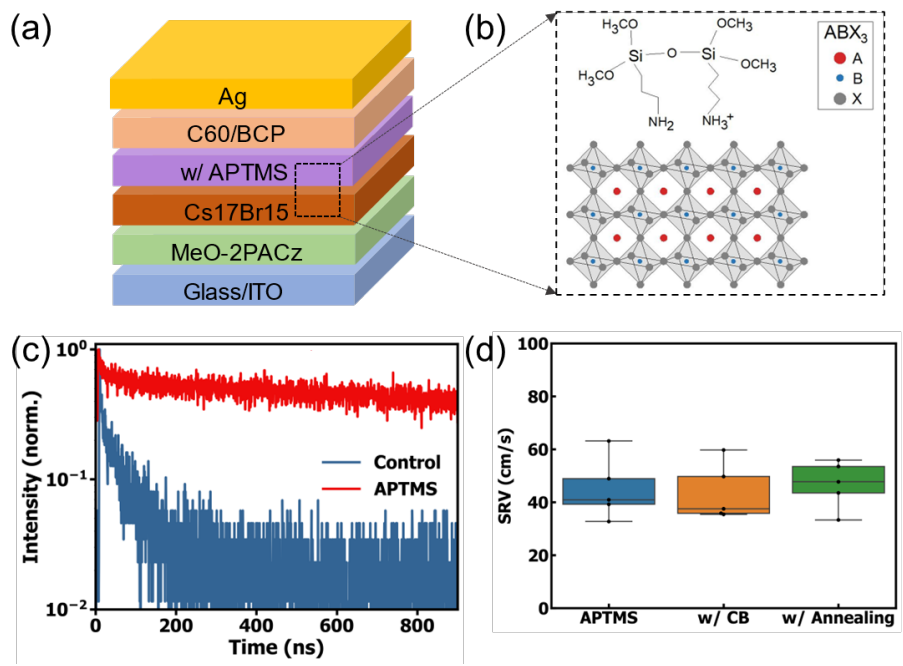

Figure 1. (a) Schematic of perovskite solar cell structure (p-i-n). (b) Schematic of passivation using APTMS molecule at the perovskite surface.(A: $FA^+,Cs^+$; B: $Pb^{2+}$; X: $I^-$, $Br^-$) (c) PL lifetime of control and APTMS-passivated perovskite films. (d) SRV of the APTMS-treated films with CB treatment and annealing.

Having shown that the passivation effect of APTMS can be sustained even after solvent washing, elevated temperature, and exposure to high vacuum, we next prepare a series of half-stack and full-stack devices (the structure of the full-stack solar cells is shown in Figure 1a). We compared the PL emission (Figure S6) of these half-stack or full-stack devices with and without APTMS passivation. We found that the passivated devices display 2 orders of magnitude higher PL emission intensity than the control stacks (Figure S6b).When in contact with a C60 ETL, the PL intensity of both passivated and control stacks is quenched due to fast interfacial recombination processes in contact with the fullerene, consistent with previous reports.[27] However, in the presence of APTMS, this fullerene-induced quenching is diminished (Figure S6).

We measured the photoluminescence quantum yield (PLQY) using an integrating sphere under 1 Sun equivalent excitation density conditions and calculate the QFLS of these stacks based on equation 1:[28-30]

$$\Delta E_F = \Delta E_{F,max} - kT|ln(PLQY_{ext})| \qquad \text{Equation 1}$$

where the $\Delta E_{F,max}$ is the Shockley-Queisser theoretical QFLS limit, k is Boltzmann constant, T is the absolute temperature, and $PLQY_{ext}$ is the measured external PLQY under 1 Sun conditions. Figure 2 shows the QFLS values for the partial device stacks at different stages of fabrication (see also Table S1). The APTMS-passivated perovskite solar cells exhibited larger QFLS than the control perovskite solar cells, indicating that a higher $V_{OC}$ could potentially be achieved for perovskite solar cells using APTMS surface passivation. The Cs17Br15 perovskite, with a bandgap of 1.63 eV, possesses a theoretical maximum QFLS of 1.35 eV. However, the as-fabricated Cs17Br15 control perovskite films on glass showed a QFLS of 1.23 eV, and the control perovskite solar cell devices showed an even smaller QFLS of 1.10 eV. We attribute such reduction in QFLS from film to solar cells to additional surface recombination at the perovskite/ETL interface, consistent with literature reports.[12] Notably, the APTMS-passivated perovskite solar cells showed a QFLS of 1.23 eV, significantly higher than that of the control perovskite solar cells. The $V_{OC}$ of perovskite solar cells is smaller than the measured QFLS under the same AM1.5 illumination intensity by ~130 mV. We attribute this remaining difference between measured $V_{OC}$ and the calculated QFLS in the perovskite solar cells to energy misalignment between the charge transporting layers and perovskite layer, suggesting there is still room to optimize energy level alignment in this particular device stack.[31] In addition to the Cs17Br15 medium bandgap perovskite, we extended the APTMS passivation to a wide bandgap mixed-cation mixed-halide perovskite (FA$_{0.83}$Cs$_{0.17}$Pb(I$_{0.75}$Br$_{0.25}$)$_3$ (denoted as Cs17Br25) with a bandgap of ~1.69 eV, which can be used for building tandem solar cells. As shown in Figure S7, the QFLS of Cs17Br25 can be increased to ~1.25 eV (Figure S7b). This increase in the QFLS of a wide bandgap perovskite is beneficial for the perovskite/silicon tandem photovoltaic devices.

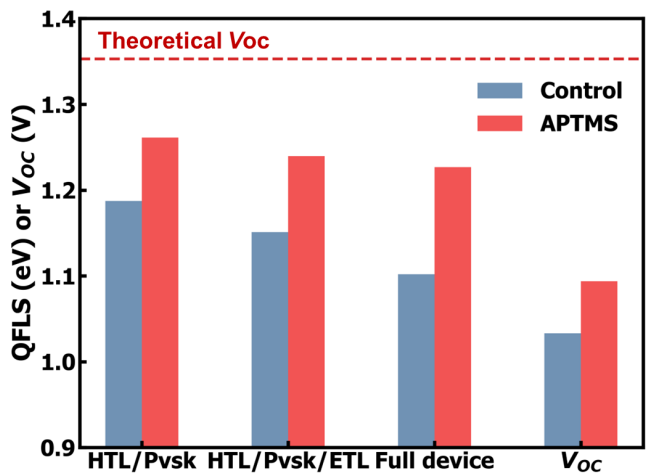

Figure 2. QFLS and measured $V_{OC}$ of the perovskite stack at different levels of perovskite solar cell device completion and operation (Pvsk: perovskite).

Thus far we have interpreted the improvements in PL lifetime, SRV and PLQY as resulting from a reduction in the surface trap density. We next verify these interpretations using excitation-correlation photoluminescence (ECPL) spectroscopy on perovskite solar cells. ECPL spectroscopy is an optical method which reveals the nonlinear interactions among the photogenerated carriers and provides insights on nonradiative decay channels of semiconducting materials. ECPL uses two temporally-delayed pump pulses with identical intensities and energy (2.638 eV).[32,33] Kanada *et al.* applied ECPL to study the recombination and trapping dynamics of carriers within metal halide perovskite materials and were able to quantitatively extract the trap density, the bimolecular recombination coefficient, and the Auger recombination rate.[32] Here, we apply ECPL spectroscopy to understand recombination and trapping dynamics of the photogenerated carriers in our control and APTMS-passivated perovskite solar cells. A schematic of the ECPL set-up can be found in the references.[33] Figure 3 shows the ECPL spectroscopy results on the control and APTMS-treated perovskite solar cells with different laser excitation fluences ranging from 4 to 36 µJ/cm$^2$ per pulse. Trap-assisted recombination results in a positive nonlinear response while Auger recombination results in a negative response. At the lowest fluence, both devices show positive ECPL values that matches with a dominating trap-limited behavior; further interpretation of ECPL signals can be found in reference 32.[32] With increasing fluence, responses of both control and APTMS-passivated perovskite solar cells decrease due to trap filling, resulting in Auger recombination becoming the dominant pathway. The APTMS-passivated perovskite solar cell shows lower initial signal compared to the control perovskite solar cell at a fluence of 4 µJ/cm$^2$, as shown in Figure 3a and Figure 3b, which we ascribe to the lower defect density in the APTMS-passivated perovskite solar cell.[32] This interpretation is in good agreement with the enhanced PL emission from this stack at low fluence. As summarized in Figure 3b and d, the control solar cell requires a high fluence of ~14.2 µJ/cm$^2$ to reach the Auger recombination regime, while the APTMS -passivated solar cell device shows signs of Auger recombination at a lower fluence of only ~8.2 µJ/cm$^2$. The lower fluence required to observe Auger recombination in the passivated solar cell is consistent with an increase in carrier lifetimes and increase in carrier density at a given excitation fluence. As such the ECPL measurement on the solar cell further confirms that APTMS can suppress the trap-assisted non-radiative recombination, which is consistent with our PL emission and QFLS results.

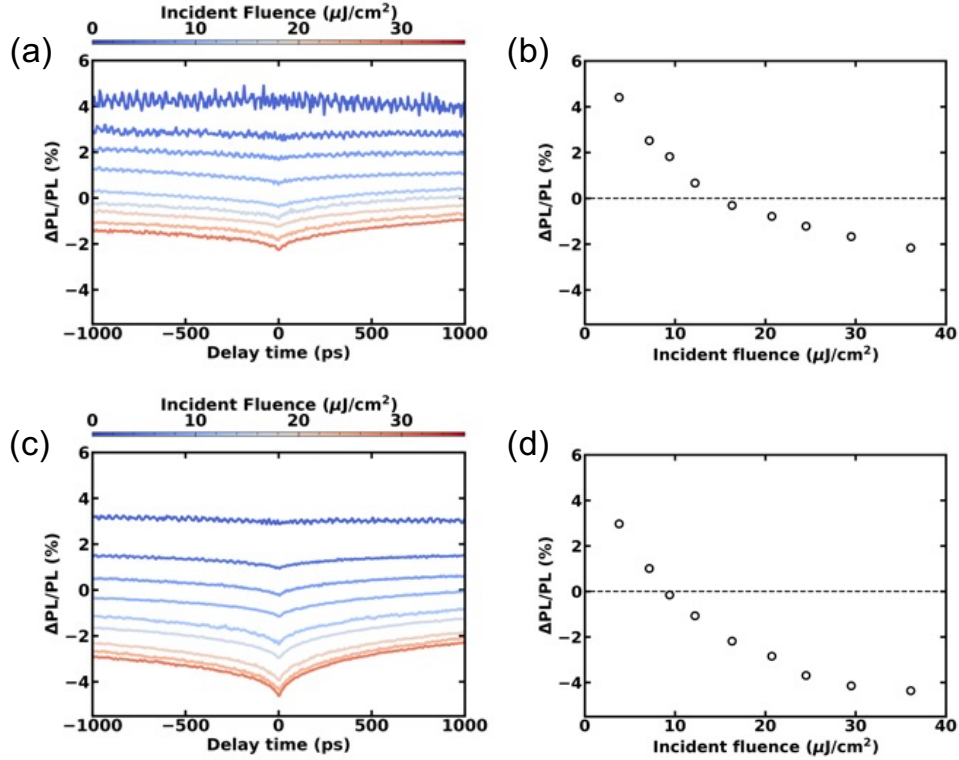

Figure 3. ECPL dynamics of control (a) and (c) APTMS-passivated perovskite solar cells. ECPL at t = 0 ps for (b) for control and (d) APTMS-passivated perovskite solar cells under various laser fluences.

The reduced SRV and enhanced QFLS in perovskite solar cells with APTMS passivation are promising indicators that point to improved device performance relative to the controls. To corroborate the effect of APTMS passivation, we fabricated solar cells with a p-i-n structure with and without APTMS surface treatment (device structure shown in Figure 1a). As expected, the $V_{OC}$ is clearly improved after the APTMS passivation as depicted in the current density-voltage (J-V) curves of the representative Cs17Br15 control device and APTMS-passivated perovskite solar cells (Figure 4a). Figure 4b shows the maximum power point tracked efficiency ($\eta_{mpp}$) for the control and APTMS-passivated perovskite solar cells. The control solar cells achieved a $\eta_{mpp}$ of

~16% while the APTMS-passivated solar cells achieved a higher $\eta_{mpp}$ of ~18%. Device performance is plotted in Figure 4c-f with statistics (based on the reverse scans) summarized in Table S2. The control solar cells displayed an average PCE of 15.90% with an FF of 72.83% and a $V_{OC}$ of 1.03 V (average), while the PCE of the APTMS passivated solar cells increased to an average value of ~18.03% with enhanced $V_{OC}$ (1.09 V) and FF (~77.78%). Overall, the APTMS-passivated solar cells displayed a consistent enhancement in $V_{OC}$ and PCE. We attribute the improvement in the device $V_{OC}$ with APTMS surface treatment to the reduced defect density, smaller SRV at the interfaces and a consequently higher QFLS. We note that in addition to the expected increase in $V_{OC}$ due to the reduction in SRV. We also observe an enhancement of the FF following APTMS passivation. Simulations show that while FF is less sensitive to SRV than $V_{OC}$, it still can benefit from reducing SRV over the range we measure in these device stacks (~125 cm/s).[15] It is possible that APTMS also helps improve FF by suppressing physical shunting and pinholes, though this cannot be distinguished from reduction in effective recombination using the current data.[34]

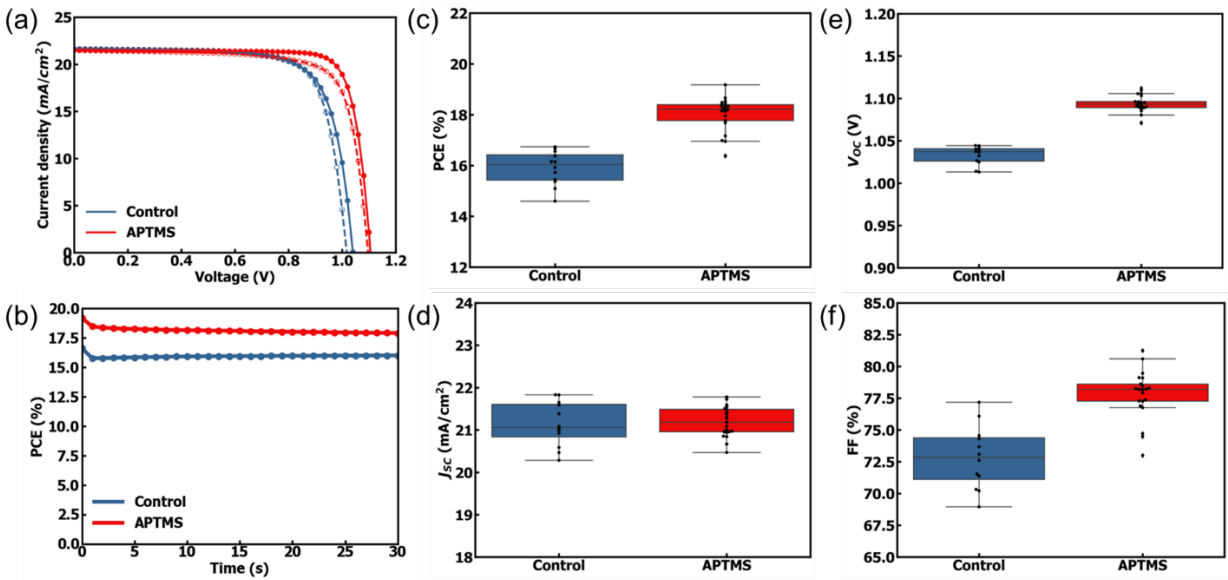

**Figure 4**. (a) *J-V* curve of representative control and APTMS-passivated solar cells (the solid and dotted lines are reverse and forward scans, respectively), (b) $\eta_{mpp}$ of control and APTMS-passivated solar cells. Statistics of (c) PCE, (d) $J_{SC}$, (e) $V_{OC}$ and (f) FF of device performance without and with APTMS passivation based on reverse scans.

In some respects, the improved device performance is surprising: one might expect APTMS, an insulating molecule, to block photocurrent extraction, yet short-circuit current ($J_{SC}$) values are comparable in both type of perovskite solar cells. Therefore, we performed a combination of scanning probe microscopy (SPM) techniques to better understand both the uniformity in coverage of the APTMS coating as well as how charge carriers are extracted from the solar cell devices. First, we used atomic force microscopy (AFM) to determine the surface coverage of the APTMS treatment by mapping the elastic modulus *E* and adhesion $F_a$, along with the topography of the passivated films since perovskite films and APTMS surfaces have anticipated significantly different mechanical properties.[35] For example, the elastic modulus of a similar siloxane polymer is ~0.1-10 MPa,[36] while those for a metal-halide perovskites are in the range 1-20 GPa.[37,38,39] Figure S8 shows topography and mechanical mapping of APTMS-treated perovskite films. The hazy part in the topography has a modulus of ~10 MPa, similar to the value expected for a siloxane polymer. Therefore, we assign these regions to the thicker areas of polymerized APTMS. The remaining area displays a much higher modulus, however, since the tip used is only sensitive to the softer polymerized APTMS, these areas are therefore indicative of the lack of APTMS, or exposed perovskites. In addition to modulus, adhesion is also obtained which is more sensitive to the surface properties than modulus measurement. While Figure S8c is well correlated with the topography results, the regions of higher adhesion are well-correlated with regions of APTMS.

These results provide insight into the reason charges are still able to be extracted in a solar cell device with APTMS passivation. With non-uniform surface coverage, some regions of APTMS are thinner allowing for charge carrier transport (presumably, improved deposition methods could be used to further optimize the coating thickness and uniformity as desired).

We also performed conductive AFM (cAFM) on half-stack devices (ITO/MeO-2PACz) to confirm how charge carriers are extracted in APTMS solar cells. Figure 5 shows the cAFM results on the control and APTMS-passivated half stacks under dark. Notably, while the APTMS-passivated half stack device shows that the current is blocked in thicker APTMS areas, we also see that there is a sufficient area of the thinner APTMS coating to allow current to be extracted. As shown in Figure 5, a 2 min APTMS layer deposition reduces the number of conducting pixels by roughly 50% while a control device is completely conductive, showing the insulating nature of APTMS. However, after deposition of the $C_{60}$ ETL, cAFM reveals uniform conductivity and photoconductivity, indicating that the smaller percentage of contact points in the APTMS-passivated device is sufficient to achieve spatially uniform current extraction (as homogenized by the fullerene layer, Figure 5d and Figure S9b). When we use an APTMS deposition time of 5 min, as we used for the best performing perovskite solar cell, the cAFM data shows only ~5% (Figure 5c) of the pixels are conducting. Nevertheless, the device data indicate that these devices still achieve a $J_{SC}$ value of ~21 mA/cm$^2$. While initially such an observation might seem surprising, these data are consistent with previous results in our group using laser beam-induced current (LBIC),[40] where we showed that the cAFM heterogeneity in perovskites is drastically reduced upon evaporation of the top contact. In this sense, it is possible for APTMS-treatment both to passivate the surface, and to reduce the interaction of the perovskite with the PL-quenching fullerene layer while still allowing sufficient contact to extract almost all the photocurrent. We speculate that similar mechanisms underpin the widespread use of insulating interlayers or dispersants such as PMMA in the literature: their primary role may be to reduce contact with PL-quenching extracting contacts, much like a passivated-emitter rear collector (PERC) silicon cell reduces surface recombination with an insulating passivating layer that has a sufficient density of holes for current extraction.[41,42]

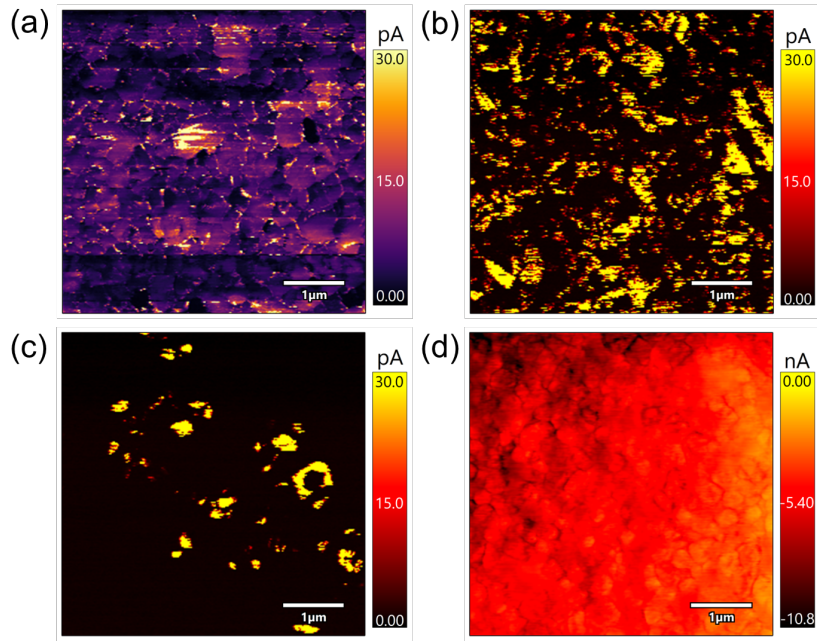

Figure 5. cAFM of (a) ITO/MeO-2PACz/Cs17Br15, 1.5 V bias, (b) ITO/MeO-2PACz/Cs17Br15/APTMS (2min) and (c) ITO/MeO-2PACz/Cs17Br15/APTMS (5min), in the dark with a bias of 1.5 V. (d) ITO/MeO-2PACz/Cs17Br15/APTMS (5min) /C60, under light (488 nm) with a bias of -1.5V. These data show that while the APTMS layer reduces the number of conducting pixels in (b) and (c), the number of contact points is sufficient to achieve uniform optoelectronic functionality once in contact with an ETL (d).

**Conclusion**

In summary, we have demonstrated a $V_{OC}$ enhancement for a mixed-cation mixed-halide perovskite solar cells by using APTMS as a surface passivator. The increase in $V_{OC}$ can be ascribed to the decrease in the defect density as verified by the enhanced PL emission, smaller SRV, and increased QFLS after APTMS passivation. In addition, we demonstrate that APTMS passivation can survive the solvent treatment, high temperature annealing and high vacuum conditions which makes it a suitable passivator for many different device fabrication or manufacturing strategies. ECPL studies further confirm that the nonradiative recombination is suppressed via APTMS passivation. Furthermore, SPM techniques reveal that the APTMS polymerizes heterogeneously at the perovskite surface but still ensure efficient extraction of carriers. This work demonstrates the importance of surface passivation in reducing defect density and improving the device performance in mixed-cation mixed-halide perovskite solar cells, while providing a potential route forward for large-scale manufacturing of improved perovskite semiconductor interfaces.

**Notes**

The authors declare no competing financial interest.

**Acknowledgements**

This work is supported primarily by the U.S. Department of Energy's Office of Energy Efficiency and Renewable Energy (EERE) under the Solar Energy Technology Office (SETO), Award Number DE-EE0008747. Atomic force microscopy imaging work is supported by the U.S. Department of Energy, Office of Basic Energy Sciences, Division of Materials Sciences and Engineering under Award DOE-SC0013957. Part of this work was carried out at the Molecular Analysis Facility, a National Nanotechnology Coordinated Infrastructure site at the University of

Washington which is supported in part by the National Science Foundation (NNCI-1542101), the Molecular Engineering & Sciences Institute, and the Clean Energy Institute. Y.S. thanks Prof. Seth R. Marder and Stephen Barlow for their suggestions and review of this work and acknowledges the use of facilities and instrumentation supported by the U.S. National Science Foundation through the UW Molecular Engineering Materials Center (MEM-C), a Material Research Science and Engineering Center (DMR-1719797). Y.S. acknowledges the financial support from the state of Washington through the University of Washington Clean Energy Institute. D.S.G. acknowledges salary and infrastructure support from the Washington Research Foundation, the Alvin L. and Verla R. Kwiram endowment, and the B. Seymour Rabinovitch Endowment.

Supporting Information Available: experimental details, XRD, PL emissions, FLIM, and AFM data